\documentclass[aps,prb,twocolumn]{revtex4}
\usepackage{graphicx}
\usepackage{dcolumn}
\usepackage{amsmath}
\usepackage{amssymb}
\usepackage{array}
\usepackage{calc}

\begin{document}

\title{Raman scattering study of Spin-Density-Wave order and electron-phonon coupling in Ba(Fe$_{1-x}$Co$_x$)$_2$As$_2$}
\author{L. Chauvi\`ere}
\author{Y. Gallais}
\author{M. Cazayous}
\author{M.A. M\'easson}
\author{A. Sacuto}
\affiliation{Laboratoire Mat\'eriaux et Ph\'enom\`enes Quantiques, UMR 7162 CNRS, Universit\'e Paris Diderot, B$\hat{a}$t. Condorcet 75205 Paris Cedex 13, France}
\author{D. Colson}
\author{A. Forget}
\affiliation{CEA Saclay, IRAMIS, Service de Physique de l'Etat Condens\'e (SPEC CNRS URA 2464), F-91191 Gif-sur-Yvette, France}

\begin{abstract}

We report Raman scattering measurements on iron-pnictide superconductor Ba(Fe$_{1-x}$Co$_x$)$_2$As$_2$ single crystals with varying cobalt $x$ content. The electronic Raman continuum shows a strong spectral weight redistribution upon entering the magnetic phase induced by the opening of the Spin Density Wave (SDW) gap. It displays two spectral features that weaken with doping, which are assigned to two SDW induced electronic transitions. Raman symmetry arguments are discussed to identify the origin of these electronic transitions in terms of orbital ordering in the magnetic phase. Our data do not seem consistent with an orbital ordering scenario and advocate for a more conventional band-folding picture with two types of electronic transitions in the SDW state, a high energy transition between two anti-crossed SDW bands and a lower energy transition involving a folded band that do not anti-cross in the SDW state. The latter transition could be linked to the presence of Dirac cones in the electronic dispersion of the magnetic state. The spectra in the SDW state also show significant coupling between the arsenide optical phonon and the electronic continuum. The symmetry dependence of the arsenide phonon intensity indicates a strong in-plane anisotropy of the dielectric susceptibility in the magnetic state.

\end{abstract}

\maketitle

\section{INTRODUCTION}

\par
The discovery of high temperature superconductivity in iron-based compounds has triggered an intense research effort to gain insight into their superconducting mechanism. The origin of electron pairing into singlet Cooper pairs in the superconducting (SC) state \cite{Zhang, Terasaki} does not seem to be explained by electron-phonon coupling alone \cite{Boeri2008}. The proximity of magnetism in these systems then plays a crucial role. The general phase diagram of iron-based superconductors \cite{Lumsden} illustrates such an interplay between magnetic and superconducting orders. Non SC parent compounds exhibit a magnetic phase transition, and superconductivity only appears with doping as magnetism weakens. In addition, some iron-based materials, particularly the Ba(Fe$_{1-x}$Co$_x$)$_2$As$_2$ compound or Co-doped Ba-122, present coexistence at the atomic scale of both orders \cite{Laplace}. It is then essential to identify the electronic structure of the magnetic phase to better understand its interplay with superconductivity.

\par
The magnetic phase is characterized by an antiferromagnetic long-range order associated with the formation of a Spin Density Wave (SDW). Magnetic moments of iron order along Fe-Fe bonds in a stripe-like pattern \cite{Huang}. In the itinerant picture, considered as a good starting point to describe these metallic systems \cite{Johnston}, a magnetic instability with the nesting vector $\overrightarrow{Q}_{SDW}$ between hole-like and electron-like Fermi surfaces induces a folding of the bands which leads to a partial reconstruction of the Fermi surfaces and the opening of a gap at the Fermi level. The nesting between circular hole pockets and ellipsoidal electron pockets is however imperfect and should lead to an anisotropic SDW gap \cite{Fernandes}. In particular, it was shown in Ba-122 that the partial destruction of electron pockets leaves small Fermi surface pockets along their elongated parts which show a linear dispersion with Dirac cone shape \cite{Richard}. Infra-red (IR) conductivity presents a spectral weight transfer from low to high energy across the magneto-structural transition for 122 compounds, consistent with the opening of a SDW gap at the Fermi level resulting from the anti-crossing of folded bands.

\par
However, it was also pointed out that a simple folding picture may not be sufficient to describe the significant reconstruction of the electronic structure in the magnetic state. Significant band shifting also occurs across magnetic ordering as reported by Angle Resolved Photo-Emission Spectroscopy (ARPES) measurements \cite{Yi2009, Yi, Fuglsang}. It was proposed that orbital ordering may imply a strong dichotomy between d$_{xz}$ and d$_{yz}$ orbital occupation. Recent ARPES measurements indeed show a lifting of d$_{xz}$ and d$_{yz}$ orbital degeneracy \cite{Yi,Lv} which could also explain the strong in-plane anisotropy observed in transport measurements \cite{Chu}. However not all the bands are clearly resolved by ARPES, which only probes the occupied state of the electronic structure making difficult the determination of the SDW gap with this technique. In addition, informations on the doping dependence of the SDW gap and the Fermi surface reconstruction associated with magnetic ordering remain quite scarce. The situation remains therefore unsettled and it is still unclear which of the scenario, orbital ordering or nesting is the driving mechanism of the magneto-structural transition.

\par
Here we report Raman scattering measurements on Co doped Ba-122 single crystals as a function of temperature and doping across the magnetic transition. We observe a systematic loss of low energy excitations across the magnetic transition due to the opening of the Spin Density Wave gap. This signature of Fermi surface reconstruction differs depending on probed symmetries. In (x'y') configuration, at x = 0, we observe two spectral features upon entering the SDW state : a low energy step-like depletion at 400 cm$^{-1}$ and a peak at 900 cm$^{-1}$, which is absent in (LL) and (xy) configurations. Upon doping the peak observed in (x'y') configuration disappears, the spectral weight depletion weakens and moves towards lower energies. Overall our data are consistent with the presence of two types of electronic transitions in the SDW state. A high energy transition between folded anti-crossed SDW bands becomes quenched upon doping. A lower energy transition involves a folded band that do not anti-cross in the SDW state. The latter transition possibly involves transition across the Dirac cone observed in ARPES measurements. The symmetry dependence of the transitions, on the other hand, seems inconsistent with a simple orbital ordering scenario. 

\par
The arsenide phonon mode across the magneto-structural transition is also studied in details because it shows a distinctive Fano lineshape evidencing strong coupling to the electronic excitations. Its behavior across the magneto-structural transition suggests a strong in-plane anisotropy of dielectric susceptibility tensor that is directly linked to the anisotropy of the electronic degrees of freedom in the SDW phase.

\section{EXPERIMENTAL DETAILS}

\par
We have measured the double layered single crystals Ba(Fe$_{1-x}$Co$_x$)$_2$As$_2$ for three different cobalt contents x = 0, 0.02, 0.045. Its phase diagram has been determined by different experimental techniques \cite{Colson, Ni, Lester, Olariu} and presents three phase transitions : a structural one at T$_s$, from a tetragonal symmetry at high temperature to an orthorhombic symmetry at low temperature, a magnetic one at T$_{\textsc{N}}$ and a superconducting one at T$_c$. For x = 0, the structural and magnetic transitions are simultaneous at T$_{s/\textsc{N}}$ = 138K. For x = 0.02, they split (T$_s$ = 120K, T$_{\textsc{n}}$ = 105K). Crystals with x = 0.045 doping (T$_s$ = 80K, T$_{\textsc{n}}$ = 64K) present superconductivity at T$_c$ = 12K. The magnetic transition disappears around optimal doping x = 0.065 (T$_c$ = 25K).

\par
Single crystals of Ba(Fe$_{1-x}$Co$_x$)$_2$As$_2$ were grown from a self-flux method by high-temperature solid-state reactions as described elsewhere \cite{Colson}. Crystals from the same batches were also studied by transport \cite{Colson}, Nuclear Magnetic Resonance (NMR) \cite{Laplace}, M$\ddot{o}$ssbauer spectroscopy \cite{Bonville} and ARPES \cite{Brouet} measurements. Typical crystal sizes are $2\times2\times0.1~$mm$^3$. Freshly cleaved single crystals were held in a vacuum of $10^{-6}~$mbar and cooled by a closed-cycle refrigerator. The spectra reported here were measured using the 514.52~nm line of an Argon-Krypton laser. Reported temperatures take into account the estimated laser heating \cite{Chauviere2009}. The scattered light was analyzed by a triple grating spectrometer (JY - T64000) equipped with a liquid nitrogen cooled CCD detector. Spectra are all corrected for the instrumental spectral response. The Raman response was extracted from the measured intensity using $\chi''\sim(1+n(\omega,T))^{-1} \times I$, where $n(\omega,T)$ is the Bose factor, $\chi''$ the imaginary part of the Raman response and $I$ the measured Raman intensity. By combining different incident and scattered photon polarizations, different Raman symmetries are probed. We now dedicate a part of this paper to clarify the notations.

\begin{figure*}
	\centering
	\includegraphics[width=0.75\linewidth]{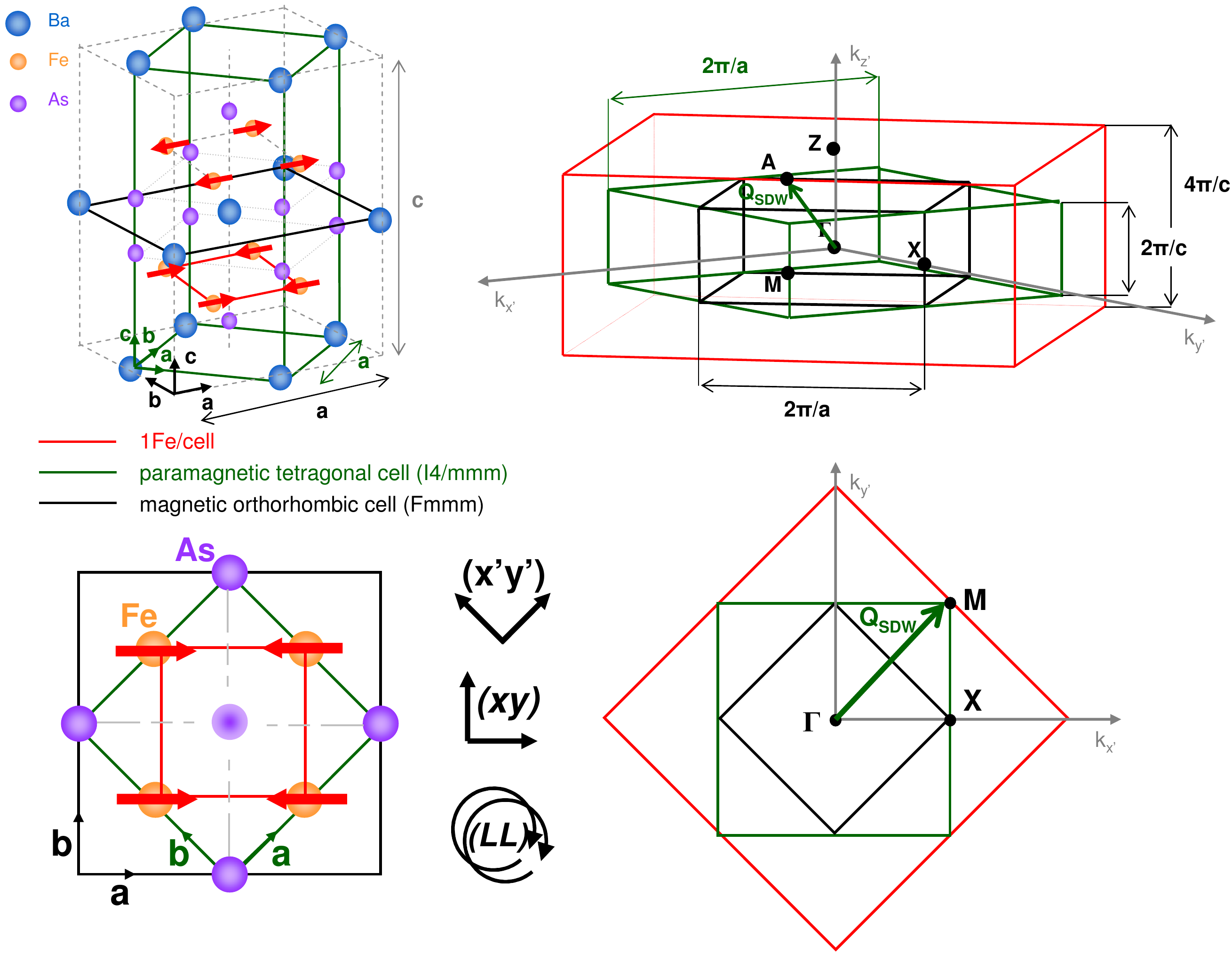}
	\caption{Different crystal cells used to describe Ba-122, with 1Fe/cell (red), paramagnetic tetragonal (green), magnetic orthorhombic (black) and their projection in the iron atoms plane. Associated first Brillouin zones and their projection in the (k$_x$, k$_y$) plane - Different configurations of incident and scattered light polarizations called (x'y') for crossed polarizations along Fe-As bonds, (xy) for crossed polarizations along Fe-Fe bonds and (LL) for circular polarizations.}
	\label{notations}
\end{figure*}

\par
Different crystallographic cells are used to describe the Ba-122 crystal structure. The simplest tetragonal crystallographic cell is based on the Fe-square plane. It contains one Fe atom per unit cell (in red in Fig.\ref{notations}). When arsenide atoms, alternating above and below Fe-planes, are taken into account, the unit cell is the `body-centered' one, which contains 2Fe/cell and is $c/2$ high \cite{Park}. The doubling of the unit cell implies a folding of the Brillouin Zone (BZ) in the reciprocal space. A simpler non magnetic tetragonal cell with 4Fe/cell, a height of $c$ and its axes (x',y') along the Fe-As bonds can also be used (in green in Fig.\ref{notations}). When the ordering of iron magnetic moments is taken into account, the larger magnetic orthorhombic cell may be used (in black in Fig.\ref{notations}) with its axes (x,y) along the Fe-Fe bonds. The first Brillouin zones, associated with these different cells, and their axes are defined in Fig.\ref{notations}. $\Gamma (0,0,0)$ is the center of the BZ and $M (\pi,\pi,0)$ the corner of the BZ, in unity of (1/a, 1/a, 1/c), in the tetragonal representation where x' and y' axes are along the Fe-As bonds. The nesting vector $\overrightarrow{Q}_{SDW}$, which creates an SDW instability is $(\pi, \pi, 2\pi)$.

\par
We used different light polarization configurations, noted (x'y') for crossed polarizations along Fe-As bonds, (xy) for crossed polarizations along Fe-Fe bonds and (LL) for left-left circular polarizations (Fig.\ref{notations}), which respectively probe the B$_{2g}$, B$_{1g}$ and A$_{1g}$ symmetries in the tetragonal phase at high temperature (\textit{I4/mmm}) \cite{Tinkham}. In the orthorhombic phase (\textit{Fmmm}), the B$_{2g}$ and A$_{1g}$ representations change to the A$_g$ representation. The Raman response in each configuration is proportionnal to the square of the projection of the Raman tensor by the light polarizations. The resulting projection transforms like $x^2-y^2$ for (x'y') configuration, $x^2+y^2$ for (LL) configuration and $xy$ for (xy) configuration, in both the tetragonal and orthorhombic phases. Additional measurements were also performed with incident and scattered light polarizations parallel and along the Fe-As bonds. This (x'x') configuration probes both A$_{1g}$ and B$_{1g}$ symmetries with equal weights (A$_{1g}$ + B$_{1g}$ in the tetragonal phase). The incident linear polarization had also a finite projection along c-axis and a small E$_g$ component, less than 10\%, was probed when using crossed polarizations. However, we compared temperature dependences of the Raman intensity with and without probing the E$_g$ component and the overall shape of the spectra were similar except for a small leak of the E$_g$(Fe, As) phonon modes.

\section{SPIN DENSITY WAVE GAP}

	\subsection{Opening of the SDW gap}

\begin{figure}
	\centering
	\includegraphics[width=1\linewidth]{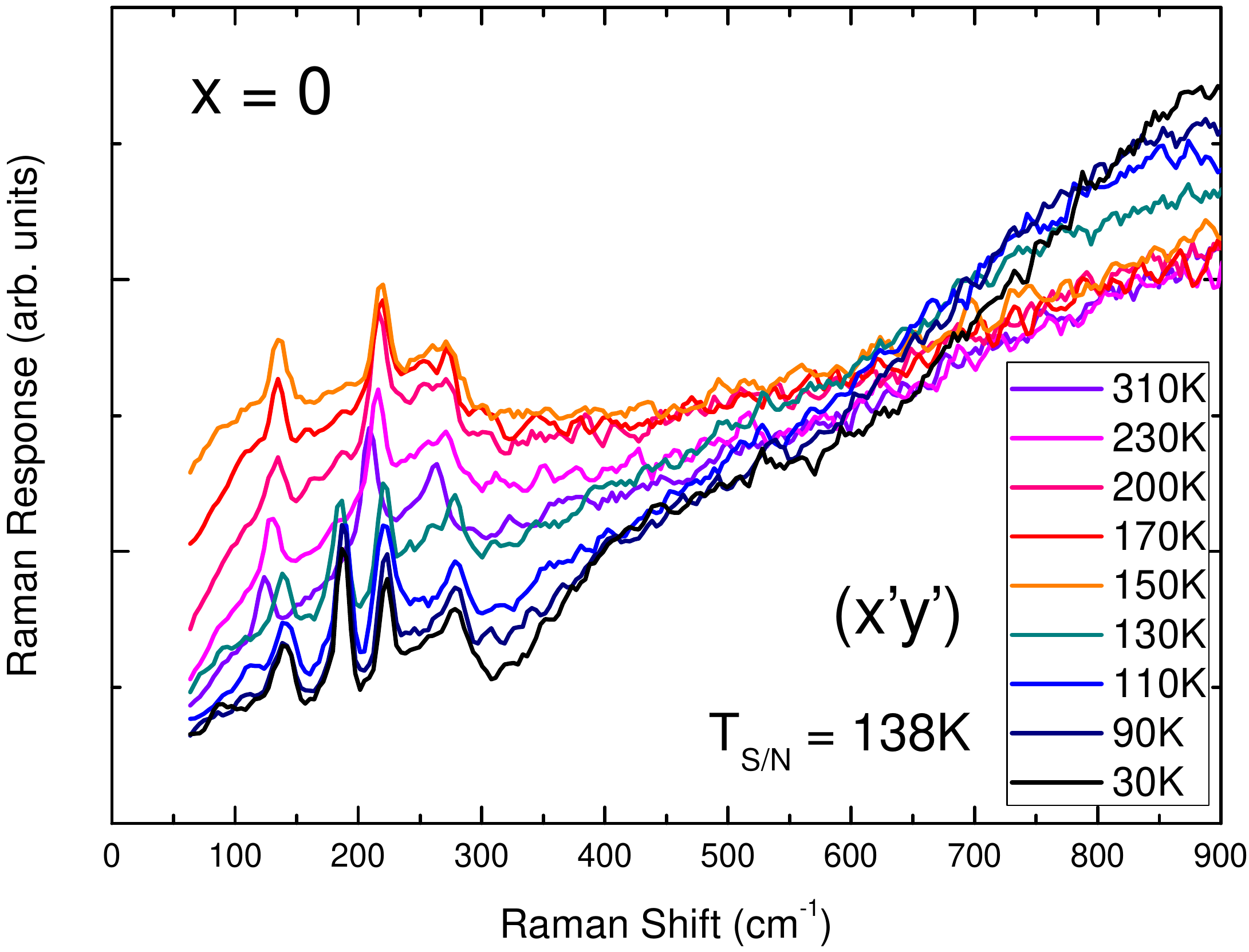}
	\caption{Temperature dependence of Raman response for the undoped compound BaFe$_2$As$_2$ in the (x'y') configuration, showing a spectral weight transfer from low to high energy across the magneto-structural transition.}
	\label{gapsdw}
\end{figure}

\par	
The temperature dependence of the Raman response for the undoped compound BaFe$_2$As$_2$ in the (x'y') configuration up to 900 cm$^{-1}$ ($\sim$110 meV) is shown in Fig.\ref{gapsdw}. Sharp peaks associated to zone-centered optical phonons can be observed. The phonon mode at around 210 cm$^{-1}$ is the B$_{1g}$(Fe) mode involving iron atoms vibrations along c-axis. It remains a small contribution here because of a small misorientation of the sample. Phonons at around 130 and 260 cm$^{-1}$ involve iron-arsenide atoms vibrations in the ab-plane. The splitting of these E$_g$(Fe, As) modes, previously reported \cite{Chauviere2009}, is not resolved here because of the lower resolution used (12~cm$^{-1}$). These phonon anomalies were already studied and are not the main interest of the present report \cite{Chauviere2009}. On the contrary, the arsenide phonon mode at around 180 cm$^{-1}$ involving arsenide atoms vibrations along c-axis becomes active in this configuration below the transition and will be discussed later in this paper.

\par
The phonon modes sit on a Raman continuum corresponding to electron-hole pairs excitations across the Fermi level. Just above the transition, we observed an increase of the response at low energy. As reported previously \cite{Chauviere2010}, the increase is systematically seen in (x'y') configuration close to the magneto-structural transition for various Co dopings. This quasi-elastic peak contribution might be associated to magnetic energy fluctuations close to the phase transition \cite{Chauviere2010}. Just below the transition, the quasi-elastic peak disappears and there is a significant loss of excitations at low energy accompanied by a spectral weight transfer from low to high energy. The spectral weight redistribution, which manifests itself the loss of low energy excitations toward higher electronic transitions, is structured around a isosbestic point at around 600-650 cm$^{-1}$ similar to what is seen in optical measurements \cite{Nakajima}, and can be assigned to the opening of a Spin Density Wave gap. We should stress here that while this behavior is qualitatively expected for the opening of a SDW gap, there is no simple formal sum-rule associated to the Raman response as in the case of optical conductivity \cite{Freericks}. This is due to the fact that the Raman response is an effective rather than true density-density response function.

	\subsection{Polarization and doping dependences}

\par
\begin{figure*}
	\centering
	\includegraphics[width=1\linewidth]{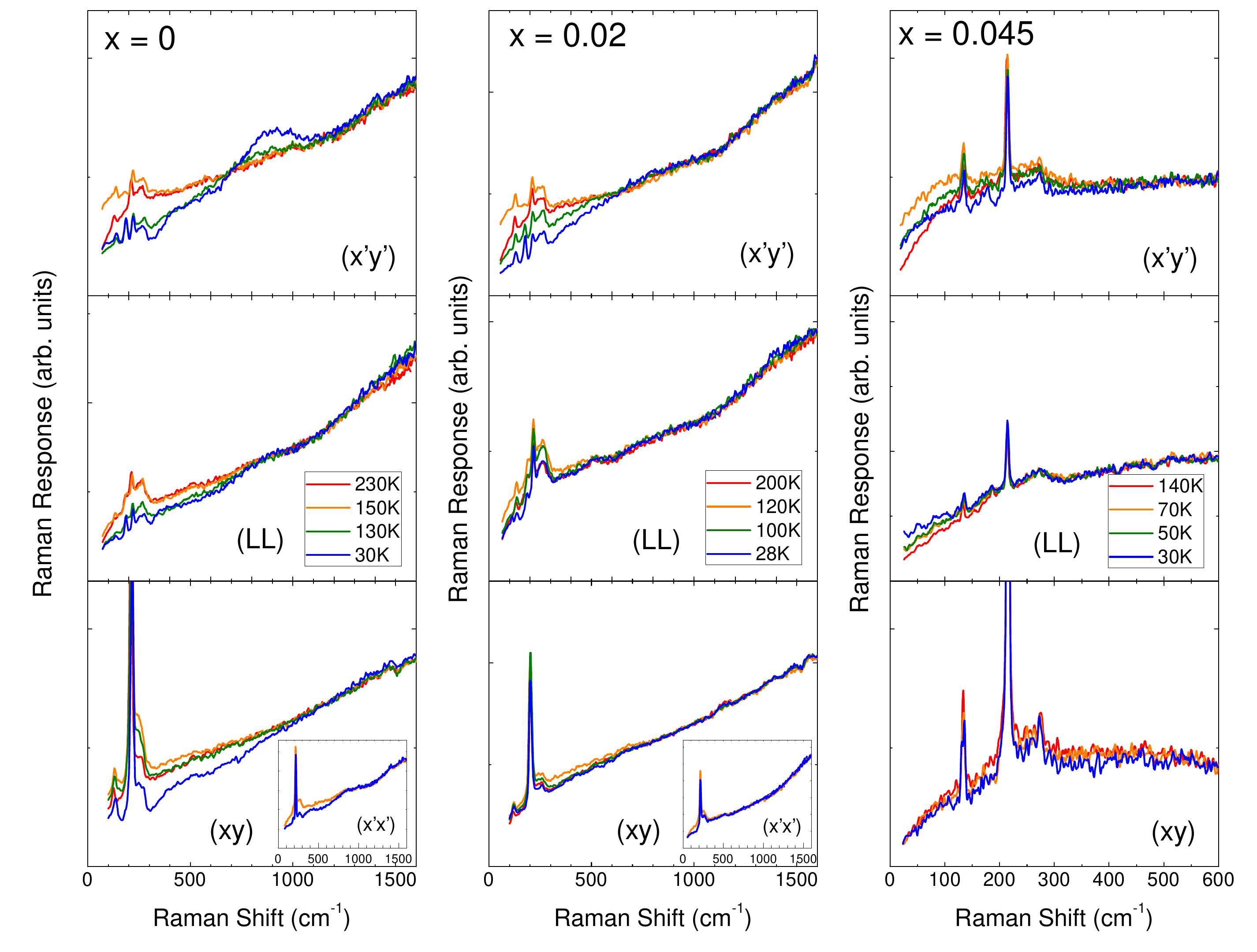}
	\caption{Raman response for x = 0, 0.02, 0.045 and for indicated configurations, at high temperature, just above the magnetic transition, just below and at low temperature (resp. red, orange, green and blue curves). It shows a spectral weight loss at low energy across the magnetic transition in all symmetries which strongly weakens with doping.}
	\label{sym}
\end{figure*}

To gain insight into the SDW gap structure, its symmetry and doping dependences over a larger energy range were studied. We measured Raman responses for four different temperatures, at high temperature, just above the magnetic transition, just below and at low temperature (respectively red, orange, green and blue curves in Fig.\ref{sym}). Special care was taken in keeping the exact same experimental conditions during a temperature dependent experiment. Spectra overlapped well and we normalized the spectra at high energy with factors less than 10\% of the signal. We performed experiments on a large energy scale, up to 1600 cm$^{-1}$ (i.e 200 meV), except for x = 0.045 doping where a significant temperature dependence was only observed below 600 cm$^{-1}$.

\par
For BaFe$_2$As$_2$, a loss of spectral weight at low energy across the magnetic transition is seen in all configurations, meaning in all symmetries (Fig.\ref{sym}). The depletion is accompanied by a peak at around 900 cm$^{-1}$ that appears only in (x'y') configuration. Upon doping, the depletion weakens gradually and moves to lower energy and the (x'y') peak essentially disappears. For x=0.045 a weak depletion is observed only in the (x'y') configuration.

\begin{figure*}
	\centering
	\includegraphics[width=1\linewidth]{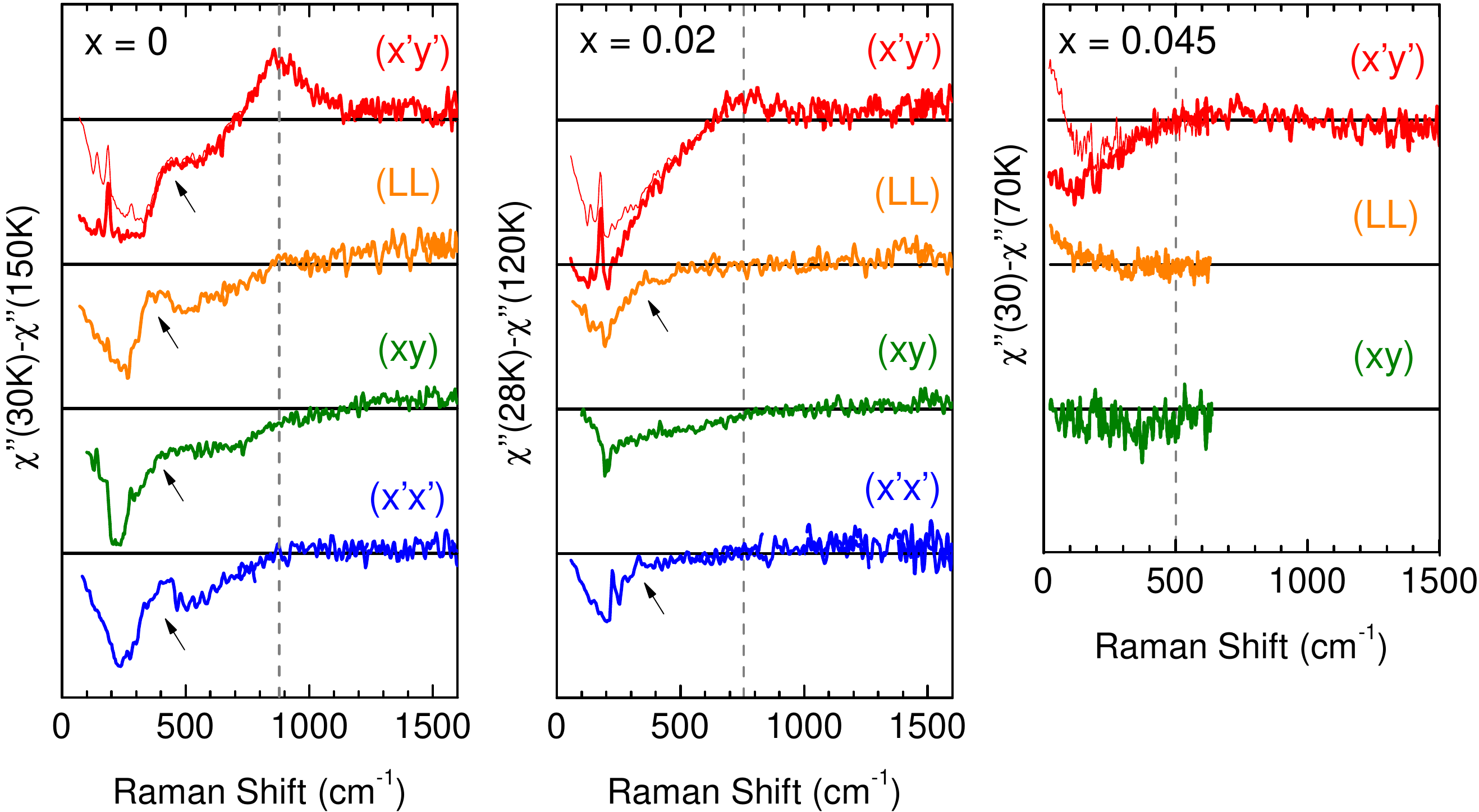}
	\caption{Subtraction between Raman responses below (T $\sim$ 30K) and above the magnetic transition (T = 150, 120 and 70K for x = 0, 0.02, 0.045), for indicated configurations and for x = 0, 0.02, 0.045. The thiner lines for (x'y') configuration are the substraction with higher temperatures references (T = 230, 200 and 140K for x = 0, 0.02, 0.045). The gray dashed lines indicate roughly the onset of the depletion and the peak position in (x'y') configuration for x = 0. The arrows indicate the low-energy step-like feature.}
	\label{sub}
\end{figure*}

\par
To highlight the impact of the magnetic transition on the Raman response $\chi$'', we have plotted the subtraction between $\chi$'' at low temperature, below the magnetic transition (T = 30K), and $\chi$'' at high temperature, just above the transition (T = 150, 120 and 70K for x = 0, 0.02, 0.045) in Fig.\ref{sub}. As mentioned before, there is a quasi-elastic contribution at low energy in (x'y') configuration above the magneto-structural transition \cite{Chauviere2010} that affects spectra below approximatively 300 cm$^{-1}$. To offset the impact of the quasi-elastic peak on the subtraction in (x'y') configuration, we have also plotted the substraction of $\chi$'' with higher temperature spectra as references (T = 230, 200K and 140K for x = 0, 0.02, 0.045) for which the quasi-elastic contribution is negligible.

\par
Figure \ref{sub} summarizes the result of the subtractions and shows the impact of the SDW gap opening on the electronic Raman response for x = 0, 0.02, and 0.045 dopings, for (x'y'), (xy) and (LL) configurations. Relative changes observed in the Raman response across the magnetic transition depend on symmetry and doping. For x~=~0, the Raman response presents two features. First, at around 900 cm$^{-1}$, the (x'y') response displays a peak while a depletion onset is observed in the other configurations around the same energy (marked by the gray dashed line in Fig.\ref{sub}). Second, there is step-like suppression at around 400 cm$^{-1}$, which is present in all configurations (marked by arrows in Fig.\ref{sub}). These two spectral features, at 400 and 900 cm$^{-1}$, are consistent with optical conductivity data on Ba-122 which also display two components at roughly the same energies in the SDW phase \cite{Akrap, Chen, Nakajima, Hu}.

\par
As already mentioned, the overall depletion considerably weakens with doping and the high energy (x'y') peak has essentially disappeared for x = 0.02 and x = 0.045. For x = 0.02, the (x'y') response still shows a sizable suppression below approximatively 750 cm$^{-1}$. The low energy step-like suppression is less pronounced but visible in (LL) configuration, between 300 and 400 cm$^{-1}$ and is hardly present in the other configurations. For x = 0.045, the effect of the SDW gap on the Raman continuum becomes weak. We only detect a small depletion in the (x'y') configuration and the spectral weight loss is gradual below 500~cm$^{-1}$.

\section{DISCUSSION}

\par{}
We have observed systematic changes in low energy electronic excitations spectra across the magnetic transition which can be associated to the opening of a SDW gap and more generally to the electronic reconstruction which occurs in the SDW phase. We will start by discussing the x = 0 spectra and its two clear spectral features shown in Fig.\ref{sdw}a : a low energy step-like depletion observed in all configurations, $\Delta_{SDW}'$, and a peak at 900 cm$^{-1}$ observed only in the (x'y') configuration, $\Delta_{SDW}$. Using Raman selection rules, we will first consider an orbital ordering scenario, which shifts bands with well defined orbital content and is expected to lift the degeneracy between $d_{xz}$ and $d_{yz}$ orbitals. We will then consider a SDW-induced band folding scenario and its associated anti-crossings, which are expected to open a SDW gap, resulting in well defined excitations at $\Delta_{SDW}$ (insert of Fig.\ref{sdw}a). Finally, we will discuss the doping dependence of the SDW gap.

	\subsection{Raman symmetries and orbitals}

\par
We first discuss the orbital content of the reconstructed bands, in the context of orbital ordering. We particularly focus on the peak at 900 cm$^{-1}$ because it shows a clear polarization dependence. Raman scattering is an inelastic process with well defined selection rules. We note $(\omega, \textbf{q})$, the energy and the momentum transferred to the system during the inelastic process. $|\textbf{q}|$ is always small with respect to the BZ extension, so that the Raman process involves essentially vertical transitions (q $\rightarrow$ 0). The scattering rate $\Gamma_{i,f}$, or electronic transition probability between an initial state $|i \rangle$ and a final state $|f \rangle$, is written : 

\begin{equation}
\Gamma_{i,f} \propto \left| \langle f| \mathcal{R}_{\textbf{q}} |i \rangle \right|^2 \delta (\epsilon_{\textbf{k}} - \epsilon_{\textbf{k-q}} - \hbar \omega)
\label{equa}
\end{equation}

where $\mathcal{R}_{\textbf{q}}$ is a Raman operator, equal to the effective electronic density operator for electronic excitations. It can be written as the projection of a Raman tensor $\widetilde{\mathcal{R}}_{\textbf{k,q}}$ by light polarizations, with \textbf{e}$_L$ and \textbf{e}$_S$ the laser (L) and the scattered (S) light polarizations respectively. 

$$ \mathcal{R}_{\textbf{q}} = \textbf{e}^*_S . (\sum_{\textbf{k}} \widetilde{\mathcal{R}}_{\textbf{k,q}}) . \textbf{e}_L $$

The Raman intensity is proportional to the sum of the scattering rate $\Gamma_{i,f}$ on all final states.

\par{}
As already stressed, the symmetry of the Raman operator $\mathcal{R}_{\textbf{q}}$ depends on polarization configurations. For (x'y') configuration, it transforms like $x^2-y^2$ (B$_{2g}$ symmetry in the tetragonal phase), for (LL) like $x^2+y^2$ (A$_{1g}$ symmetry in the tetragonal phase), and for (xy) like $xy$ (B$_{1g}$ symmetry in the tetragonal phase). It connects the initial $|i \rangle$ and final $|f \rangle$ states during the inelastic scattering process. An even operator connects states with the same parity ; an odd operator connects states with different parities. The parity of initial and final states is determined by the orbital character of electronic bands involved in the transition. Electronic structure of Ba-122 presents five bands crossing the Fermi level. All of them originate from the five \textit{d}-orbitals of iron \textit{d}$_{xy}$, \textit{d}$_{xz}$, \textit{d}$_{yz}$, \textit{d}$_{x^{2}-y^{2}}$ and \textit{d}$_{z^{2}}$  where (x,y) directions are taken along Fe-Fe bonds as is usually done in most band structure calculations \footnote{We note here that iron d-orbitals may point toward arsenide atoms, as sometimes used to define hopping parameters \cite{Ran}, but are usually pointing toward other iron atoms, as commonly used to describe ferro-orbital order \cite{Kruger, Lv}.}. As we work on twinned samples, we are not able to distinguish x and y directions. We then deduce the allowed electronic transitions in each configurations by finding the non-zero matrix elements connecting two orbitals of $xy$, $xz$, $yz$, $x^{2}-y^{2}$ or $z^{2}$ symmetry with a Raman operator of $x^2-y^2$, $x^2+ y^2$ or $xy$ symmetry. In the (x'y') configuration, $x^{2}-y^{2} \leftrightarrow z^{2}$ transition is the only allowed transition and there is no intra-orbital transition. In the (xy) configuration, $xy \leftrightarrow z^{2}$ and $xz \leftrightarrow yz$ transitions are possible but not intra-orbital ones. In the (LL) configuration, only the intra-orbital transitions are allowed.

\par{}
It immediately follows that the peak at 900~cm$^{-1}$ observed in (x'y') configuration, if interpreted in terms of a transition between bands with well defined orbital character, must correspond to electronic transitions between $\textit{d}_{x^{2}-y^{2}}$ and $\textit{d}_{z^2}$ derived bands. On the other hand, the step at around 400~cm$^{-1}$ appears in all symmetries and thus can not be assigned to any specific electronic transition between bands with well defined orbital content. 

\par{}
The orbital assignment of the (x'y') peak is inconsistent with band structure calculations where the \textit{d}$_{xz}$, \textit{d}$_{yz}$ and \textit{d}$_{xy}$ orbitals play the leading role in the electronic reconstruction across the SDW transition. In particular, electronic transitions corresponding to the same spectral features observed in optical conductivity were assigned by Dynamical Mean Field Theory (DMFT) calculations \cite{Yin} to transitions between the d$_{xz}$ and d$_{xy}$ orbitals and within the d$_{xz}$ orbital for the  900~cm$^{-1}$ peak and the 400~cm$^{-1}$ step respectively. We should also note that recent ARPES measurements display significant disagreement with calculations in the orbital assignment of the bands crossing the Femi level \cite{Zhang2011}. The reshuffling of spectral weight observed in the Raman response across the SDW transition thus does not seem to be linked with an orbital ordering which would lead to transitions between orbital related bands. This leads us to discuss the observed feature in terms of more conventional SDW induced band folding and associated anti-crossing leading to the observed spectral features.

	\subsection{SDW induced band folding}

\begin{figure}
	\centering
	\includegraphics[width=1\linewidth]{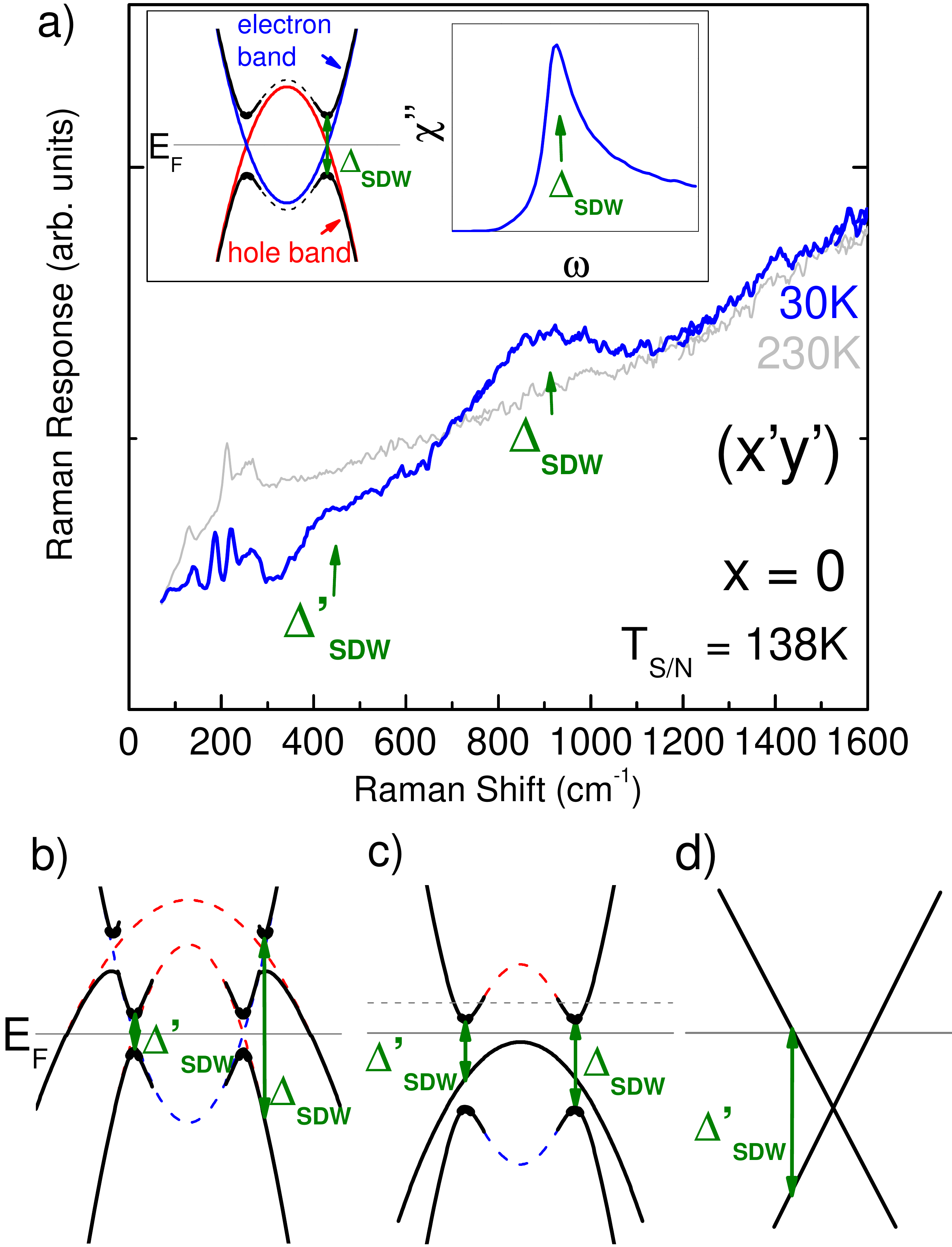}
	\caption{a) Low temperature Raman response, compared to the high temperature one, in the (x'y') configuration for the parent compound Ba-122 showing two spectral features at $\Delta_{SDW}$ and $\Delta_{SDW}'$. Insert : anti-crossing of a hole and an electron band inducing the opening of a SDW gap $\Delta_{SDW}$ and associated Raman response calculated in the case of perfect nesting \cite{Valenzuela, Vanyolos}. For simplicity constant Raman vertices along the hole and electron Fermi surfaces and a small phenomenological broadening $\gamma=0.05\Delta_{SDW}$ were taken. We note that in our notation $\Delta_{SDW}$ is a two-particle gap. In the case of perfect nesting, this gap is exactly twice the single particle gap, as extracted by ARPES experiments. b) Illustration of two anti-crossings between one electron band and two hole bands. c) Illustration of two anti-crossed bands and an additional hole band, which does not interact with the others. d) Illustration of a Dirac cone.}
	\label{sdw}
\end{figure}

\par{}
We now discuss the Raman response at low temperature for the parent compound, in terms of SDW induced band folding alone (Fig.\ref{sdw}a). We first consider two bands to explain the two features, $\Delta_{SDW}$ and $\Delta_{SDW}'$. In the case of perfect nesting, the system would become insulating in the magnetic phase. As it is not the case \cite{Hu, Chen}, the Fermi surface is partially gapped and the SDW gap is strongly anisotropic. The imperfect nesting of the two considered bands, due to different size, shape and dimensionality \cite{Vorontsov, Fernandes2010, Malaeb}, could lead to an in-plane anisotropy and a strong k$_z$ dependence of the gap. A simple two-band scenario with an anisotropic SDW gap appears however to be inconsistent with the distinct Raman symmetry dependence of the two spectral features. This scenario also seems ruled out by optical conductivity measurements showing both features for E//ab-plane and only the low energy one for E//c-axis \cite{Chen}. 

\par{}
An alternative approach to explain the two components is to consider two distinct SDW gaps, $\Delta_{SDW}$ and $\Delta_{SDW}'$, resulting from two distinct anti-crossings in k-space. One anti-crossing comes from the $\overrightarrow{Q}_{SDW}$ nesting between a hole pocket centered around $\Gamma$ and an electron pocket centered around $M$. The anti-crossing of these two bands opens a gap $\Delta_{SDW}$ in the excitation spectrum, provided that the Fermi level lies in between the SDW bands. The other spectral feature, $\Delta_{SDW}'$ could originate from the anti-crossing of a second hole-like band with the same electron band, as illustrated in Fig.\ref{sdw}b. This scenario is similar to the one applied for antiferromagnetic chromium \cite{Asano} to explain optical conductivity measurements in the SDW phase \cite{Barker, Boekelheide}.

\par{}
Two anti-crossings are however not necessarily needed to explain the two spectral features appearing in the SDW phase. We can also consider two anti-crossed bands and a third band which does not interact with the other bands in the SDW phase (Fig.\ref{sdw}c). The third band could be a hole band which, while folded in the SDW phase, does not show significant SDW mixing. This picture is similar to the one proposed by Eremin et al. \cite{Eremin} to explain the lifting of degeneracy between ($\pi$,0) and (0,$\pi$) SDW orders in the unfolded Brillouin zone (1Fe/cell). It gives rise to two vertical electronic transitions at $\Delta_{SDW}$ and $\Delta_{SDW}'$, which can be associated with the step-like and the peak features of the (x'y') response at x = 0. Moreover, because of SDW coherence factors, the spectral weight of the transitions is much stronger for momentum close to the anti-crossing points, as marked by bolder black line in Fig.\ref{sdw}. This is a consequence of the formation of the SDW condensate which involves mostly electrons and holes close to these points. Therefore, electronic transitions between two folded bands around anti-crossing points are expected to have significantly more spectral weight than transitions involving a band that is not participating in the SDW condensate. The presence of a high energy peak and a broader low energy step-like suppression thus seems to advocate for such a scenario and more generally for the presence of band(s) which do not significantly mix in the SDW phase. 

\par{}
The assignment of the step-like suppression as arising from uncoupled bands given above, could also be consistent with the presence of Dirac cones. The Dirac cones are formed by folded bands which do not open a gap in the magnetic phase. They were measured \cite{Richard, Kim, Yi} and calculated \cite{Ran, Morinari, Sugimoto} as crossing points between 20 and 30 meV below the Fermi level along the $\Gamma$-$M$ direction in the tetragonal representation. The crossing creates new vertical transitions between occupied and unoccupied states above an energy defined as twice the energy difference between the Fermi level and the Dirac point as illustrated in Fig.\ref{sdw}d \footnote{This relation is exact only for perfect untilted Dirac cones.}. The step-like depletion at 400~cm$^{-1}$, or 50 meV, would then originate from Dirac crossing point located at 25 meV below the Fermi level, in agreement with ARPES measurements \cite{Kim, Yi}. A similar scenario was proposed by Sugimoto et al. to explain optical conductivity data in the SDW phase \cite{Sugimoto}. In their calculation of the optical conductivity using a five-band Hubbard model, the step-like suppression emanates from interband transitions across the Dirac cones while the high energy feature is assigned to transitions between SDW gapped bands.

	\subsection{Doping dependence}

\par{}
The overall spectral weight redistribution related to magnetic ordering strongly weakens with doping. It is consistent with ARPES data \cite{Liu} showing a reduced portion of reconstructed Fermi surface upon cobalt doping. The two discussed SDW spectral features are strongly reduced with doping. The first step in the depletion becomes broad for x = 0.02 and is hardly present at x = 0.045. The peak accompanying the spectral weight transfer in the (x'y') configuration is suppressed. The relatively rapid suppression of SDW induced features is consistent with optical measurements on Co-Ba122 \cite{Nakajima}. The energy at which the high energy depletion disappears, marked by gray dashed lines in Fig.\ref{sub}, decreases as expected from the evolution of T$_{\textsc{n}}$ : 900$\pm50$, 750$\pm100$ and 500$\pm100$ cm$^{-1}$ for T$_{\textsc{n}}$ = 138, 105, 64K respectively.

\par
The rapid collapse of the (x'y') peak between x = 0 and x = 0.02 can be explained if we simply consider cobalt content increasing as a rigid band shift. The collapse would then result from the filling of an electron band upon doping. In the picture of Fig.\ref{sdw}c, an increase of the Fermi level shown in dashed gray line would fill unoccupied states at the anti-crossing points and lead to the disappearance of electronic transitions associated with the SDW gap opening, leaving weaker interband transitions away from the anti-crossing. Such a change in the Fermi surface topology, or Lifschitz transition, was already discussed to explain thermoelectric power and Hall coefficient measurements, showing abrupt changes between the x$\leq$0.02 and the x$\geq$0.03 data \cite{Mun, Liu}.

\par{}
To summarize, the symmetry dependence of the two spectral features does not seem to be consistent with a transition between bands with well-defined orbital content. We have then discussed the spectral weight transfer due to the magnetic ordering for the undoped compound BaFe$_2$As$_2$ in the itinerant model. The (x'y') peak most likely comes from an SDW induced anti-crossing between electron and hole-like bands whereas the low energy step-like suppression is associated to electronic transitions involving folded but most likely uncoupled bands, possibly at Dirac points. The polarization dependence of the Raman response would then be related to different k-space dependence of the Raman vertices in the band representation rather than the orbital content of the bands \cite{Mazin, Nowadnick}. This point needs further study but closely resembles the polarization dependence of the superconducting response observed at higher doping \cite{Muschler, Chauviere2010, Mazin}.

\section{ARSENIDE PHONON MODE}

\par
\begin{figure}
	\centering
	\includegraphics[width=1\linewidth]{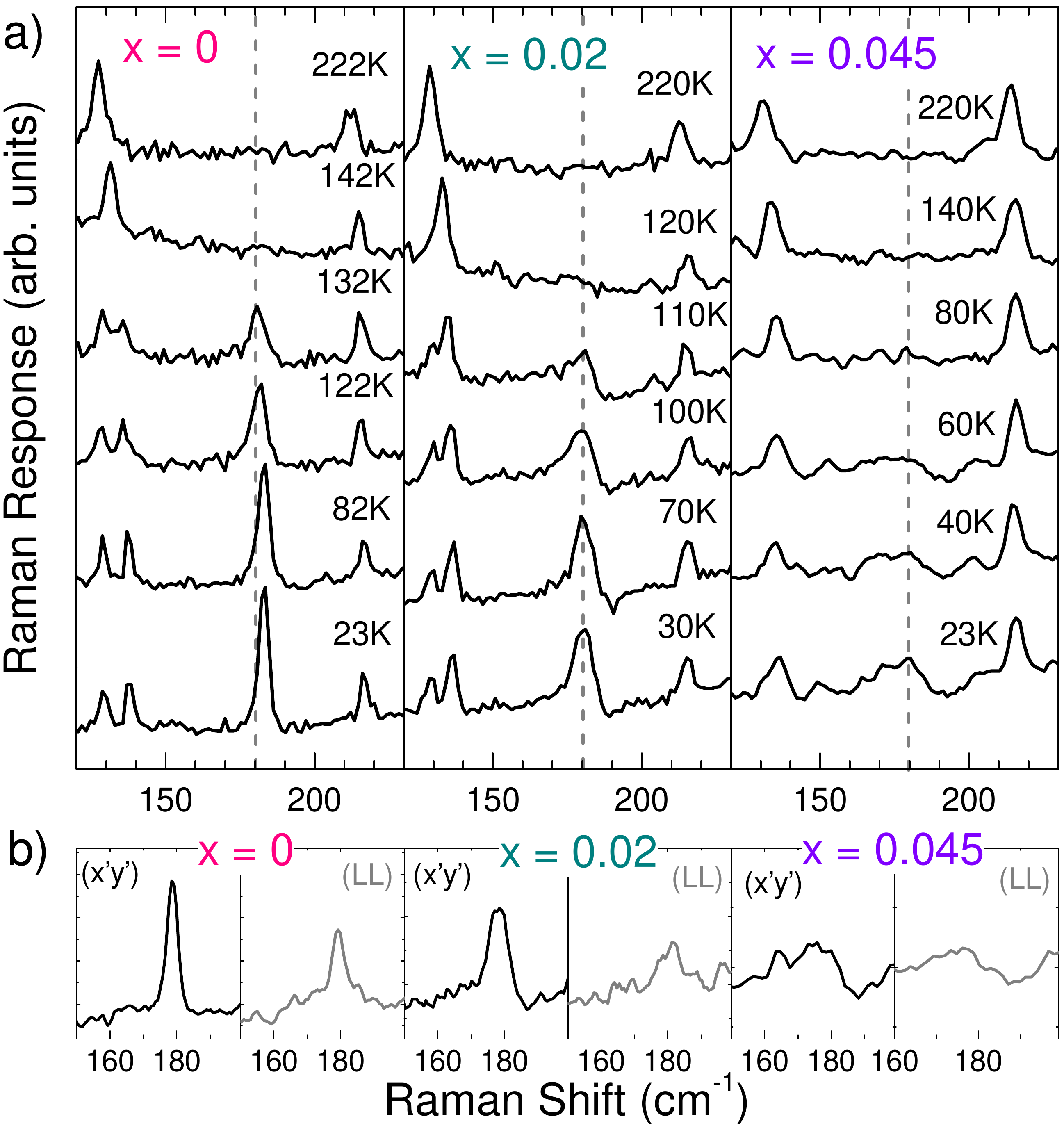}
	\caption{a) Raman Response temperature dependence in the (x'y') configuration zoomed on the As phonon mode energy at around 180~cm$^{-1}$, for x~=~0, 0.02, 0.045. Spectra were translated for clarity. The gray dashed line is at 180~cm$^{-1}$. b) Raman response at low temperature (T~$\sim$~30K) in the (x'y') and (LL) configurations for same dopings.}
	\label{arsenic}
\end{figure}

{}
In this section, we focus on the arsenide (As) phonon mode, which involves arsenide atoms vibrations along c-axis and has A$_{1g}$ symmetry in the tetragonal phase\cite{Litvinchuk}. A more detailed view of the temperature dependence of the Raman response around the As phonon energy in the (x'y') configuration for x = 0, 0.02, 0.045 is displayed in Fig.\ref{arsenic}a. The data were taken with a higher resolution (about 2 cm$^{-1}$) compared to the spectra presented in the previous section. Besides the splitting of the doubly degenerate E$_g$ mode, we observe the appearance of the As phonon mode in the orthorhombic phase at around 180~cm$^{-1}$. By comparing data in the (x'y') and (LL) configurations at low temperature (Fig.\ref{arsenic}b), we notice that the As phonon peak intensity is comparable in each configuration for all doping. Its integrated intensity is even a factor of about 1.5 larger in the (x'y') than in the (LL) configuration. While As phonon anomalies were reported for undoped 122 compounds before \cite{Choi, Chauviere2009, Rahlenbeck, Sugai}, the present data shows that the As phonon lineshape also strongly evolves with doping. For x = 0, it is sharp and symmetrical at low temperature. With doping however, it strongly broadens, its intensity weakens and its frequency decreases, as highlighted by the gray dashed line at 180~cm$^{-1}$ (Fig.\ref{arsenic}a). It also displays a distinctive asymmetrical lineshape below the magneto-structural transition.

\par{}
We first discuss the As phonon mode appearance in the (x'y') configuration at low temperature. As reported in Eq.\ref{equa}, the Raman intensity is proportional to the square modulus of the Raman tensor projected by incoming and outgoing light polarizations (\textbf{e}$_L$,\textbf{e}$_S$). Raman tensor irreducible representations for the Raman phonon modes are reported in Table \ref{tab} in both the tetragonal and the orthorhombic phases. They are written with (x',y') axes along Fe-As bonds in the tetragonal structure and with (x,y) axes along Fe-Fe bonds in the orthorhombic one. The 2E$_g$ (\textit{I4/mmm}), the B$_{2g}$ and B$_{3g}$ (\textit{Fmmm}) symmetries have been omitted and there is no Raman active phonon of B$_{2g}$ symmetry in the tetragonal phase \cite{Litvinchuk}. Raman tensor projections for the As phonon mode and for different light polarizations configurations (\textbf{e}$_L$,\textbf{e}$_S$) are reported in Table \ref{conf} for both phases.

\begin{table}
\centering

\scalebox{1.1}{
\begin{tabular}{ccc}
\hline \noalign{\smallskip}
\multicolumn{3}{c}{Tetragonal phase \textit{I4/mmm}} \\
$A_{1g}$ & $B_{2g}$ & $B_{1g}$ \\
$ \begin{pmatrix} a & 0 & 0 \\ 0 & a & 0 \\ 0 & 0 & c \end{pmatrix}$
&
$ \begin{pmatrix} 0 & d & 0 \\ d & 0 & 0 \\ 0 & 0 & 0 \end{pmatrix}$
&
$ \begin{pmatrix} e & 0 & 0 \\ 0 & -e & 0 \\ 0 & 0 & 0 \end{pmatrix}$ \\
\noalign{\smallskip} \hline \noalign{\smallskip}
\multicolumn{3}{c}{Orthorhombic phase \textit{Fmmm}} \\
\multicolumn{2}{c}{$A_{g}$} & $B_{1g}$ \\
\multicolumn{2}{c}{$ \begin{pmatrix} a' & 0 & 0 \\ 0 & b' & 0 \\ 0 & 0 & c \end{pmatrix}$}
&
$ \begin{pmatrix} 0 & e & 0 \\ e & 0 & 0 \\ 0 & 0 & 0 \end{pmatrix}$ \\
\noalign{\smallskip}
\hline

\end{tabular}}
\caption{Raman tensor irreducible representations, with (x',y') axes along Fe-As bonds in the tetragonal structure (\textit{I4/mmm}) and with (x,y) axes along Fe-Fe bonds in the orthorhombic one (\textit{Fmmm}).}
\label{tab}
\end{table}

\par{}
Classically the phonon Raman tensor is the first derivative of dielectric susceptibility tensor over the phonon induced atomic displacement \cite{book}. Electrical field of the incident light induces an electrical polarization in the crystal described by the dielectric susceptibility tensor $\epsilon_{ij}$, with $i$ and $j$ extending over the three space coordinates. In the linear response approximation, Raman response is obtained from vibration-induced polarization, a'~=~$\partial \epsilon_{xx} / \partial u$ and b'~=~$\partial \epsilon_{yy} / \partial u$, with $u$ the atomic displacements associated with the As phonon mode. It attributes the sensibility of in-plane electronic polarizability to lattice fluctuations along $c$-axis.

\begin{table}
\centering
\scalebox{1.3}{
\begin{tabular}{c|c|c}
 & \includegraphics[width=0.1\linewidth]{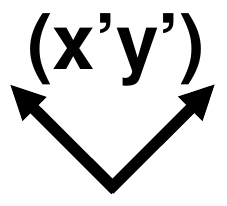} & \includegraphics[width=0.1\linewidth]{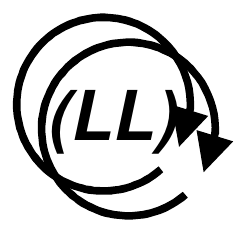} \\
\hline
\textit{I4/mmm} & - & a \\
\hline
\textit{Fmmm} & (a'-b')/2 & (a'+b')/2 \\
\end{tabular}}

\caption{Different light polarization configurations (\textbf{e}$_L$, \textbf{e}$_S$) named (x'y') for crossed polarizations along Fe-As bonds (first column) and (LL) for circular polarizations (second column), and their associated As phonon Raman tensor projection in the tetragonal high temperature and orthorhombic low temperature phases (first and second rows).}
\label{conf}
\end{table}

\par
As shown in Table \ref{conf}, the As phonon becomes active in the (x'y') configuration in the orthorhombic phase because the orthorhombic x and y axes are inequivalent. Its intensity in the (x'y') configuration is proportional to $|\frac{\partial \epsilon_{xx}}{\partial u} - \frac{\partial \epsilon_{yy}}{\partial u}|^2$ while it is proportional to $|\frac{\partial \epsilon_{xx}}{\partial u} + \frac{\partial \epsilon_{yy}}{\partial u}|^2$ in the (LL) configuration. From the observed relative intensities between both configurations, we therefore have $|\frac{\partial \epsilon_{xx}}{\partial u} - \frac{\partial \epsilon_{yy}}{\partial u}|^2 \geq |\frac{\partial \epsilon_{xx}}{\partial u} + \frac{\partial \epsilon_{yy}}{\partial u}|^2$. We can conclude that the diagonal Raman tensor elements of the As phonon mode, $\frac{\partial \epsilon_{xx}}{\partial u}$ and $\frac{\partial \epsilon_{yy}}{\partial u}$, are different by at least a factor of 10 and could even have opposite signs \footnote{As we work on twinned samples, we can not distinguish between the two in-plane directions (x,y). We consequently cannot extract which Raman tensor element is particularly weak or positive.}. Such in-plane anisotropy cannot be easily explained by anisotropic lattice constants alone. The orthorhombic distortion, characterized by the ratio of in-plane lattice constants, has a maximum of 0.36\% for x~=~0 and decreases monotonically with increasing cobalt concentration \cite{Prozorov}. The small orthorhombic distorsion would suggest $\frac{\partial \epsilon_{xx}}{\partial u} \sim \frac{\partial \epsilon_{yy}}{\partial u}$ and thus vanishingly small intensity of the As phonon in the (x'y') configuration compared to the (LL) configuration. The strikingly large in-plane anisotropy of the susceptibility derivative suggests that the coupling between electrons and the As phonon is strongly anisotropic in the (x,y) plane in the SDW phase. 

\par{}
The electronic in-plane anisotropy of Co doped Ba-122 properties was recently illustrated by resistivity measurements performed on detwinned samples \cite{Chu}. They showed that Ba(Fe$_{1-x}$Co$_x$)$_2$As$_2$ compounds develop a large anisotropy around the magneto-structural transition, with the resistivity along the shorter b-axis being greater than along a-axis. We show here that this electronic anisotropy is also reflected in the lattice degrees of freedom.

\begin{figure}
	\centering
	\includegraphics[width=1\linewidth]{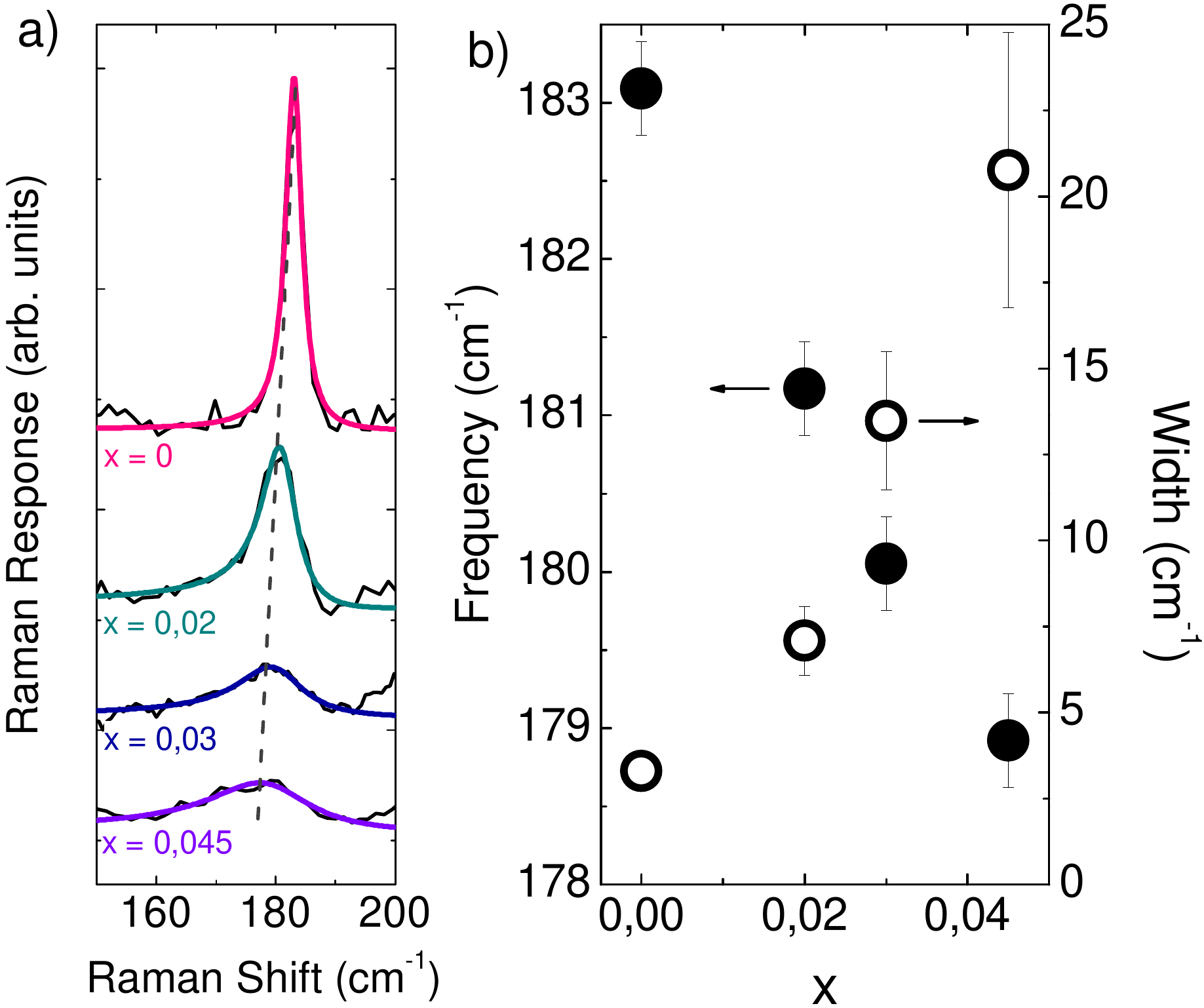}
	\caption{a) Raman response at low temperature (T = 23/30K) for x = 0, 0.02, 0.03, 0.045 dopings in the (x'y') configuration, and their associated fit with a Fano line shape (see text). The spectra were translated vertically for clarity. b) As phonon frequency (full circles) and width (empty circles) as a function of doping.}
	\label{fit}
\end{figure}

\par{}
Strong coupling between As phonon and electronic degrees of freedom also affects its lineshape. Contrary to the other Raman active phonons which remain symmetric, the As phonon is clearly asymmetric for x = 0.02 and x = 0.045, particularly just below the structural transition temperature. The asymmetry is the signature of the interaction between a discrete (phonon) state and a broad (electronic) continuum. The interference between a discrete level and a continuum leads to a Fano lineshape asymmetry \cite{Klein}. The Raman response for a Fano lineshape phonon is written as: 

$$ \chi''(\omega) \propto \frac{1}{\frac{\gamma}{2}q^2} \frac{(q+\alpha(\omega))^2}{1+\alpha(\omega)^2}~,~~~~\alpha(\omega) = \frac{\omega-\omega_c}{\gamma/2} $$

where q defines the asymmetry, $\omega_c$ the phonon frequency and $\gamma$ the phonon linewidth. $\gamma$ is proportionnal to both the electronic density of state and the electron-phonon coupling constant squared. $\omega_c$ can be written as $\omega_c = \omega^0_c - \Delta\omega_c$, where $\omega^0_c$ is the resonance frequency of the phonon without electron-phonon coupling and $\Delta\omega_c$ the shift of frequency induced by electron-phonon coupling, which is proportionnal to both the Hilbert transform of the electronic density of state and the electron-phonon coupling constant squared.

\par{}
In Fig.\ref{fit}a, the low temperature Raman responses and their associated Fano lineshape fits are plotted for different dopings. The extracted q parameter is negative as the anti-resonance is at higher energy than the phonon peak energy and the symmetric Lorentzian lineshape is recovered in the q$\rightarrow \infty$ limit. For x = 0, the As phonon lineshape presents a weak asymmetry ($|q|~>$~30) and tends to a Lorentzian lineshape. For x~=~0.02, 0.03 and 0.045, the low temperature asymmetries are significantly larger ($|q|$~$\approx$~6.5). 

\par{}
The low temperature As phonon frequency and linewidth extracted from Fano lineshape fits are reported in Fig.\ref{fit}b. Their temperature dependences were already reported for undoped Ba-122 and Sr-122 \cite{Choi, Rahlenbeck}. The As phonon mode strongly broadens with doping, as its extracted low temperature linewidth increases from 2 cm$^{-1}$ (i.e. resolution limited) to 9~cm$^{-1}$ between x~=~0 and 0.045. A significant softening of the phonon frequency with doping is also observed, from 183 to 178.5~cm$^{-1}$ between x~=~0 and x=0.045. The softening cannot be accounted by structural effects only, as the c lattice constant decreases by only 0.3\% between 0 and 0.045 doping \cite{Ni} which would lead to a softening \cite{Chauviere2009} of the order of 0.8 cm$^{-1}$. A similar anomalous softening of the As phonon mode was reported in SmFeAsO upon F doping and attributed to strong spin-phonon coupling and its calculated frequency was shown to be strongly dependent on the Fe magnetic moment both in 1111 and 122 systems \cite{LeTacon, Zbiri, Reznik}.

\par{}
In an electron-phonon coupling picture, as suggested by the Fano lineshape asymmetry, the doping induced strong broadening and frequency shift can be linked to the increase of the electronic density of state at the Fermi level upon doping because of a combination of reduced SDW gap amplitude and changes in the Fermi topology (or Lifschitz transition) as discussed in the previous section. Similarly, for x = 0 and x = 0.02, the As phonon asymmetry is larger around the transition and significantly decreases upon further cooling, resulting in more symmetric lineshapes at low temperature (Fig.\ref{arsenic}). This behavior can be linked to the strong decrease of the density of state at low energy due to the opening of the SDW gap below T$_{\textsc{n}}$.

\par
Combined with the unusual symmetry dependence of its intensity in the SDW phase, the evolution of the As phonon lineshape with both temperature and doping reported here highlights its extreme sensitivity to the underlying electronic and magnetic degrees of freedom \cite{Kuroki2009}. Our results agree with recent DFT calculations which show that in addition to conventional spin-phonon coupling effects which shift the phonon frequencies, the electron-phonon coupling constant is enhanced in the magnetic phase \cite{Boeri, Yndurain}. These effects were shown to be particularly strong for the out of plane As vibrations discussed here. This strong electron-phonon coupling calls for a reassessment of its role in the understanding of the phase diagram \cite{Saito}.

\section{CONCLUSION}

\par
We have reported Raman scattering measurements on Co doped Ba-122 single crystals as a function of temperature and doping. The electronic Raman continuum displays clear signatures of the opening of a doping dependent SDW gap. For undoped Ba-122, signatures of the Fermi surface reconstruction differs according to probed symmetries and are stronger in (x'y') configuration where two clear features are observed : a low energy step-like suppression and a high energy peak. Upon doping the peak observed in (x'y') configuration disappears and the signatures of Fermi surface reconstruction weaken while moving towards lower energies. The symmetry dependence of the Raman response appears inconsistent with a simple orbital ordering scenario and advocates for a more conventional band-folding picture with two types of electronic transitions in the SDW state, a high energy transition between folded anti-crossed SDW bands and a lower energy transition involving a folded band that does not anti-cross in the SDW state. The latter transition possibly involves transition across the Dirac cones observed in ARPES.

\par
The SDW transition has a strong impact on the arsenide phonon optical mode. A distinctive Fano lineshape is observed just below the transition demonstrating a strong coupling between the As phonon and the electronic continuum. This is in agreement with theoretical works showing enhanced electron-phonon coupling for the out of plane As phonon in the magnetic phase. The unusual symmetry dependence of the As mode in the SDW phase also suggests a strong in-plane anisotropy of the dielectric susceptibility tensor that is directly linked to the anisotropy of the electronic degrees of freedom in the SDW phase reported in transport measurements on detwinned crystals.

\par
We thank Thomas Devereaux and Gabriel Kotliar for useful discussions. Financial support from the Agence National de la Recherche (ANR 'Pnictides') is acknowledged.

\bibliography{ref}

\end{document}